\documentclass[aps,nofootinbib,commonaddress,preprint,showpacs]{revtex4}
\usepackage{amsmath}
\usepackage{amssymb}
\usepackage{graphicx}
\usepackage{color}

\begin{document}

\title{Dispersion and damping of potential surface waves in a degenerate plasma}
\author{Yu.~Tyshetskiy}
\email{y.tyshetskiy@physics.usyd.edu.au}
\author{D.~Williamson}
\author{R.~Kompaneets}
\author{S.V.~Vladimirov}
\affiliation{School of Physics, The University of Sydney, NSW 2006, Australia}

\date{\today}
\received{}
%\revised{}

\begin{abstract}
Potential (electrostatic) surface waves in plasma half-space with degenerate electrons are studied using the quasi-classical mean-field kinetic model. The wave spectrum and the collisionless damping rate are obtained numerically for a wide range of wavelengths. In the limit of long wavelengths, the wave frequency $\omega$ approaches the cold-plasma limit $\omega=\omega_p/\sqrt{2}$ with $\omega_p$ being the plasma frequency, while at short wavelengths, the wave spectrum asymptotically approaches the spectrum of zero-sound mode propagating along the boundary. It is shown that the surface waves in this system remain weakly damped at all wavelengths (in contrast to strongly damped surface waves in Maxwellian electron plasmas), and the damping rate nonmonotonically depends on the wavelength, with the maximum (yet small) damping occuring for surface waves with wavelength of $\approx5\pi\lambda_{F}$, where $\lambda_{F}$ is the Thomas-Fermi length. 
\end{abstract}
%\pacs{52.35.-g, 52.65.-y, 52.25.Dg, 05.30.-d} 	%52.35.-g: Waves, oscillations, and instabilities in plasmas and intense beams
						%52.65.-y: Plasma simulation
						%52.25.Dg: Plasma kinetic equations 
						%05.30.-d: Quantum statistical mechanics

\maketitle

%\tableofcontents

\newpage

\section{Introduction \label{sec:intro}}
% \textcolor{blue}{\textit{The recent progress in nanotechnology has revived the interest in... 
% Spasers, plasmonics, etc., see ``The Case for Plasmonics'' article in Applied Physics.}}

It has long been known~\cite{Ritchie_1957,Trivelpiece_Gould_1959,Guernsey_1969} that bounded plasmas support a special type of collective electrostatic and electromagnetic excitations -- the surface plasma waves -- whose field and energy are concentrated near, and propagate along, the plasma boundaries. The surface waves (SW) in various (classical) plasmas have been extensively studied, and they have found many applications (see Ref.~\cite{Vladimirov_progress_1994} and references therein). Yet recently, due to a remarkable progress in nanotechnology, the interest in surface waves supported by various nanostructures (especially metallic structures such as thin films and tiny metallic particles), and in their interaction with light, has been revived. It is believed that light-induced surface excitations in such structures may offer a route to faster, smaller, and more efficient electronics, as well as new technology~\cite{Plasmonics_Science_2010}. In particular, one could note such recent advents in the new and promising area of quantum nanoplasmonics as the development of the concept of spaser~\cite{SPASER_PRL_2003} followed by its further development into a lasing spaser~\cite{Zheludev_etal_2008}, and the experimental demonstration of a spaser-based nanolaser~\cite{Noginov_etal_2009,Lasers_go_nano}.

In view of these developments, understanding the properties of surface waves in various metallic (and semiconductor) structures, bounded by vacuum or dielectric, is thus important. Such understanding requires using models describing the dynamic response of such structures to self-consistent electromagnetic fields, that also appropriately take into account the relevant quantum effects arising from quantum nature of free charge carriers in such structures and, in general, from their (quantum) interaction with each other and with the underlying ion lattice. These quantum effects may significantly alter the properties of the surface waves; see, e.g., Refs.~\cite{Vladimirov_Kohn_1994,Marklund_etal_new_quantum_limits}. 

Recently, the dispersion relation of surface waves in one of the basic structures modeling a nanoplasmonic device -- a semi-bounded collisionless quantum plasma with degenerate electrons -- was obtained in Ref.~\cite{Lazar_etal_2007} using the quantum fluid theory (QFT) approach~\cite{Manfredi_Haas_PRB_2001}. In the electrostatic limit, the authors of Ref.~\cite{Lazar_etal_2007} obtained for the frequency of surface waves in this structure:
\begin{equation}
\Omega = \frac{1}{\sqrt{2}}\left(1+\sqrt{\frac{3}{2}}K_\parallel\sqrt{1+3H^2K_\parallel^2}\right),  \label{eq:Lazar_disp}
\end{equation}
where $\Omega=\omega/\omega_p$, $K_\parallel=k_\parallel\lambda_F$, $H=\hbar\omega_p/2mv_F^2$, $\omega_p=(4\pi e^2 n_e/m_e)^{1/2}$ is the electron plasma frequency, $v_F=\hbar\sqrt[3]{3\pi^2 n_e}/m_e$ is the electron Fermi velocity, $\lambda_F=v_F/\sqrt{3}\omega_p$ is the Thomas-Fermi length, $\omega$ is the SW frequency, $k_\parallel$ is the SW wave vector along the plasma boundary, $e$, $m_e$ and $n_e$ are electron charge, mass and number density, respectively, and $\hbar$ is the reduced Planck constant.
However, the validity of the dispersion relation (\ref{eq:Lazar_disp}) obtained in Ref.~\cite{Lazar_etal_2007} (as well as of the similar dispersion relation obtained in Ref.~\cite{Marklund_etal_new_quantum_limits}) is limited by the validity of the QFT approach itself~\cite{Manfredi_Haas_PRB_2001,Vlad_Tysh_UFN_2011}, and is restricted only to long waves, $K_\parallel\ll 1$. Moreover, the QFT approach by its nature completely ignores the purely kinetic effect of collisionless damping of surface waves, which is known to be significant, e.g., for potential SW in warm electron plasma at short wavelengths~\cite{ABR_book}. To overcome these limitations, a kinetic approach to the problem of SW in semi-bounded plasma with degenerate electrons is needed.

In this paper, we study potential surface waves in semi-bounded collisionless quantum plasma with Fermi-degenerate electrons, using the initial value problem solution for the semi-classical Vlasov-Poisson system~\cite{Guernsey_1969}. We obtain the dispersion and collisionless damping rate of these waves, valid for both long and short wavelengths, and report on a surprising result that these waves remain weakly damped for all values of $K_\parallel$ (i.e., for all wavelengths), unlike, e.g., in plasma with Maxwellian electrons in which the SWs are weakly damped only for $K_\parallel\ll 1$ and quickly become strongly damped as $K_\parallel$ increases. We also report a nontrivial nonmonotonic dependence of the damping rate on $K_\parallel$, featuring a maximum damping at $K_\parallel\approx 0.4$ (corresponding to the surface wave length of $\approx5\pi\lambda_{F}$).

\section{Method}
\subsection{Model and Assumptions \label{sec:model}}
% \textcolor{blue}{\textit{Formulate the problem at hand: plasma, assumptions, model equations, geometry. Discuss the applicability of the collisionless approximation, and argue why only the quantum statistics is taken into account, while quantum recoil (aka quantum tunneling) is ignored.}}

We consider a uniform nonrelativistic quantum plasma consisting of mobile electrons with charge $e$, mass $m_e$ and number density at equilibrium $n_e=n_0$, and immobile uniform background of singly charged ions with number density $n_0$ that neutralizes the electron charge at equilibrium. The plasma is assumed to be confined to the region $x<0$ by a sharp perfectly reflecting wall located at $x=0$. We will be interested in evolution of an initial perturbation to the equilibrium state of the system's electronic component applied at time $t=0$. 

In general, this system of many interacting quantum particles (electrons) can be described by the quantum analog of the Bogoliubov-Born-Green-Kirkwood-Yvon (BBGKY) hierarchy~\cite{Klim_Statfizika,Balescu_1975} of equations for the electron $j$-particle quantum distribution functions $W_j(\mathbf{r}_j,\mathbf{p}_j,t)$ (also called the $j$-particle Wigner functions), where $j=1,2,...,N$, $\mathbf{r}_j$ and $\mathbf{p}_j$ are the $3j$-dimensional vectors denoting the sets of coordinates and canonical momenta of system particles, and $N$ is the total number of electrons in the system. In this hierarchy, each of the equations for the $j$-particle quantum distribution function contains the $(j+1)$-particle distribution function, making the whole set of $N$ equations coupled, and thus prohibitively large to solve. In practice, however, this hierarchy can be truncated by making a physically justified assumption about correlation of particles due to their interaction. In particular, for a system of weakly interacting particles, with a small plasma coupling parameter $Q=U_{\rm int}/\epsilon_{\rm kin}\ll 1$ (here $U_{\rm int}$ is the characteristic potential energy of particle interaction, and $\epsilon_{\rm kin}$ is the characteristic kinetic energy of plasma particles), the two- and higher-order particle correlations can be ignored, leading to the collisionless mean-field approximation~\cite{Tyshetskiy_etal_PoP_2011} involving only one equation for the one-particle quantum distribution function $W_1(\mathbf{r,p},t)$, where $\mathbf{r}$ and $\mathbf{p}$ are now the 3-dimensional vectors of particle coordinate and momentum. In the quasi-classical approximation, with the effect of quantum recoil ignored (see Sec.~\ref{sec:applicability}), this equation reduces to the Vlasov equation for the one-particle classical distribution function $f(\mathbf{r,v},t)$, where $\mathbf{v}=\mathbf{p}/m_e$ is electron velocity. For a system of particles with electrostatic interaction, in the chosen geometry, the Vlasov equation for $f(\mathbf{r,v},t)=f(x,\mathbf{r}_\parallel,v_x,\mathbf{v}_\parallel,t)$ (where $\mathbf{r}_\parallel$ and $\mathbf{v}_\parallel$ are, respectively, the components of $\mathbf{r}$ and $\mathbf{v}$ parallel to the boundary) reads
\begin{equation}
\frac{\partial f}{\partial t} + v_x\frac{\partial f}{\partial x} + \mathbf{v}_\parallel\cdot\frac{\partial f}{\partial\mathbf{r}_\parallel} - \frac{e}{m_e}\left(\frac{\partial\phi}{\partial x}\frac{\partial f}{\partial v_x} + \frac{\partial\phi}{\partial\mathbf{r}_\parallel}\cdot\frac{\partial f}{\partial\mathbf{v}_\parallel}\right) = 0, \label{eq:Vlasov}
\end{equation}
where the electrostatic potential $\phi(x,\mathbf{r}_\parallel,t)$ is defined by the Poisson's equation
\begin{equation}
-\nabla^2\phi = 4\pi e\left[\int{f(\mathbf{r,v},t)d^3\mathbf{v}} - n_0\right]. \label{eq:Poisson}
\end{equation}
In the absence of fields, the equilibrium distribution function of plasma electrons $f_0(\mathbf{v})$ is defined by the Pauli's exclusion principle for fermions, and for low electron temperatures $T_e/\epsilon_F\ll 1$ (where $T_e$ is the electron temperature, $\epsilon_F=m_ev_F^2/2=(3\pi^2\hbar^3 n_e)^{2/3}/2m_e$ is the electron Fermi energy) it becomes
\begin{equation}
f_0(v)=\begin{cases}
\frac{3 n_0}{4\pi v_F^3} = \frac{2 m_e^3}{(2\pi\hbar)^3} &\mbox{if }v\leq v_F, \\
0 &\mbox{if }v > v_F,
\end{cases}  \label{eq:f0}
\end{equation}
corresponding to fully degenerate electron distribution.

The condition of specular reflection of plasma electrons off the boundary at $x=0$ implies
\begin{equation}
f(x=0,\mathbf{r}_\parallel,-v_x,\mathbf{v}_\parallel,t) = f(x=0,\mathbf{r}_\parallel,v_x,\mathbf{v}_\parallel,t).  \label{eq:boundary}
\end{equation}
 
\subsection{Initial value problem}
% \textcolor{blue}{\textit{Formulate the initial value problem: initial perturbation and its evolution in dimensionless variables (from Guernsey). Discuss the terms in the solution.}}

We now introduce a small initial perturbation $f_p(x,\mathbf{r}_\parallel,v_x,\mathbf{v}_\parallel,t=0)$ to the equilibrium electron distribution function $f_0(v)$, $|f_p(x,\mathbf{r}_\parallel,v_x,\mathbf{v}_\parallel,t=0)|\ll f_0(v)$, and study the resulting evolution of the system's charge density $\rho(x,\mathbf{r}_\parallel,t)=e\left[\int{f(x,\mathbf{r_\parallel,v},t)d^3\mathbf{v}}-n_0\right]$, and hence of the electrostatic potential $\phi(x,\mathbf{r}_\parallel,t)$ defined by (\ref{eq:Poisson}). Introducing the dimensionless variables $\Omega=\omega/\omega_p$, $\mathbf{K}=\mathbf{k}\lambda_F$, $\mathbf{V}=\mathbf{v}/v_F$, $X=x/\lambda_F$, $\mathbf{R}_\parallel=\mathbf{r}_\parallel/\lambda_F$, $\lambda_F=v_F/\sqrt{3}\omega_p$, and following Guernsey~\cite{Guernsey_1969}, the solution of the formulated initial value problem for $\rho(X,\mathbf{R}_\parallel,T)$ with the boundary condition (\ref{eq:boundary}) is
\begin{eqnarray}
\rho(X,\mathbf{R}_\parallel,T) &=& en_0\tilde{\rho}(X,\mathbf{R}_\parallel,T), 
\end{eqnarray}
where
\begin{eqnarray}
\tilde{\rho}(X,\mathbf{R}_\parallel,T) &=& \frac{1}{(2\pi)^3}\int_{-\infty}^{+\infty}{dK_x{\ e}^{i K_x X}\int{d^2\mathbf{K}_\parallel}{\ e}^{i\mathbf{K}_\parallel\cdot\mathbf{R}_\parallel}\ \tilde{\rho}_{\mathbf{k}}(T)}, \\
\tilde{\rho}_{\mathbf{K}}(T) &=& \frac{1}{2\pi}\int_{i\sigma-\infty}^{i\sigma+\infty}{\tilde{\rho}(\Omega,\mathbf{K}){\ e}^{-i\Omega T}d\Omega},\ \text{with }\sigma>0. \label{eq:inv_Laplace} 
\end{eqnarray}
The integration in (\ref{eq:inv_Laplace}) is performed in complex $\Omega$ plane along the horizontal contour that lies in the upper half-plane ${\rm Im}(\Omega)=\sigma>0$ above all singularities of the function $\tilde{\rho}(\Omega,\mathbf{K})$. The function $\tilde{\rho}(\Omega,\mathbf{K})$, defined as the Laplace transform of $\tilde\rho_\mathbf{K}(T)$:
\begin{equation}
\tilde{\rho}(\Omega,\mathbf{K}) = \int_0^\infty{\tilde\rho_\mathbf{K}(T)\ e^{i\Omega T} dT}, \label{eq:Laplace}
\end{equation}
is given by
\begin{eqnarray} 
\tilde{\rho}(\Omega,\mathbf{K}) &=& \frac{i}{\varepsilon(\Omega,K)}\int{d^3\mathbf{V}\frac{G(\mathbf{V,K})}{\Omega-\sqrt{3} \mathbf{K\cdot V}}} \nonumber \\
&+&\frac{iK_\parallel}{2\pi\zeta(\Omega,K_\parallel)}\left[1-\frac{1}{\varepsilon(\Omega,K)}\right]\int_{-\infty}^{+\infty}{\frac{dK_x'}{{K^\prime}^2\ \varepsilon(\Omega,K')}\int{d^3\mathbf{V}\frac{G(\mathbf{V},\mathbf{K^\prime})}{\Omega-\sqrt{3}\mathbf{K'\cdot V}}}},  \label{eq:rho(w,k)} 
\end{eqnarray}
where the Fourier transforms $G(\mathbf{V,K})$ and $G(\mathbf{V,K'})$ of the (dimensionless) initial perturbation are defined by
\begin{eqnarray}
G(\mathbf{V,K}) &=& \int_{-\infty}^{+\infty}{dX {\ e}^{-i K_x X}\  \tilde{g}(X,V_x,\mathbf{V}_\parallel,\mathbf{K}_\parallel)}, \\
G(\mathbf{V,K'}) &=& \int_{-\infty}^{+\infty}{dX {\ e}^{-i K_x' X}\  \tilde{g}(X,V_x,\mathbf{V}_\parallel,\mathbf{K}_\parallel)}, 
\end{eqnarray}
with
\begin{eqnarray}
\tilde{g}(X,V_x,\mathbf{V}_\parallel,\mathbf{K}_\parallel) &=& \int{d^2\mathbf{R}_\parallel {\ e}^{-i\mathbf{K_\parallel\cdot R_\parallel}}\  \tilde{f_p}(X,\mathbf{R}_\parallel,V_x,\mathbf{V}_\parallel,0)}, \\
\tilde{f_p}(X,\mathbf{R}_\parallel,V_x,\mathbf{V}_\parallel,0) &=& \frac{v_F^3}{n_0} f_p(X,\mathbf{R}_\parallel,V_x,\mathbf{V}_\parallel,0),
\end{eqnarray}
where $K_\parallel=|\mathbf{K}_\parallel|$, $K=|\mathbf{K}|$, $\mathbf{K}=(K_x,\mathbf{K}_\parallel)$, $K'=|\mathbf{K'}|$, and $\mathbf{K'}=(K_x',\mathbf{K}_\parallel)$.
The functions $\varepsilon(\Omega,K)$ and $\zeta(\Omega,K_\parallel)$ in (\ref{eq:rho(w,k)}) are defined (for ${\rm Im}(\Omega)>0$) as follows:
\begin{eqnarray}
\varepsilon(\Omega,K) &=& 1 - \frac{1}{\sqrt{3}K^2}\int{\frac{\mathbf{K}\cdot\partial\tilde{f}_0(\mathbf{V})/\partial\mathbf{V}}{\Omega-\sqrt{3}\mathbf{K\cdot V}}d^3\mathbf{V}}, \label{eq:epsilon} \\
\zeta(\Omega,K_\parallel) &=& \frac{1}{2} + \frac{K_\parallel}{2\pi}\int_{-\infty}^{+\infty}{\frac{dK_x}{K^2\ \varepsilon(\Omega,K)}}, \label{eq:zeta}
\end{eqnarray} 
with 
\[
\tilde{f}_0(\mathbf{V}) = \frac{v_F^3}{n_0}f_0(\mathbf{V}) = \frac{v_F^3}{n_0}\left.f_0(\mathbf{v})\right|_{\mathbf{v}=v_F\mathbf{V}}. 
\]
For fully degenerate plasma with electron distribution (\ref{eq:f0}), the function $\varepsilon(\Omega,K)$ becomes~\cite{Gol'dman_1947,ABR_book}:
\begin{equation}
\varepsilon(\Omega,K) = 1 + \frac{1}{K^2}\left[1-\frac{\Omega}{2\sqrt{3}K}\ln\left(\frac{\Omega+\sqrt{3}K}{\Omega-\sqrt{3}K}\right)\right],  \label{eq:epsilon_degenerate}
\end{equation}
where $\ln(z)$ is the principal branch of the complex natural logarithm function.

Note that the solution~(\ref{eq:rho(w,k)}) differs from the corresponding solution of the transformed Vlasov-Poisson system for infinite (unbounded) uniform plasma only in the second term involving $\zeta(\Omega,K_\parallel)$; indeed, this term appears due to the boundary at $x=0$.

% \textcolor{blue}{\textit{Say that the poles $\varepsilon(\Omega,K)=0$ of $\tilde{\rho}(\Omega,\mathbf{K})$ in (\ref{eq:inv_Laplace}) define the properties of volume waves, and the poles $\zeta(\Omega,K_\parallel)=0$ of $\tilde{\rho}(\Omega,\mathbf{K})$ in (\ref{eq:inv_Laplace}) define the properties of surface waves, as shown by Guernsey. Below we discuss the analytic properties and evaluation of $\zeta(\Omega,K_\parallel)=0$, from which the dispersion and damping of the surface wave solution will be obtained.}}

The definition (\ref{eq:Laplace}) of the function $\tilde\rho(\Omega,\mathbf{K})$ of complex $\Omega$ has a sense (i.e., the integral in (\ref{eq:Laplace}) converges) only for ${\rm Im}(\Omega)>0$. Yet the long-time evolution of $\tilde{\rho}_\mathbf{k}(T)$ is obtained from (\ref{eq:inv_Laplace}) by displacing the contour of integration in complex $\Omega$ plane from the upper half-plane ${\rm Im}(\Omega)>0$ into the lower half-plane ${\rm Im}(\Omega)\leq0$~\cite{Landau_1946}. This requires the definition of $\tilde\rho(\Omega,\mathbf{K})$ to be extended to the lower half-plane, ${\rm Im}(\Omega)\leq 0$, by analytic continuation of (\ref{eq:rho(w,k)}) from ${\rm Im}(\Omega)>0$ to ${\rm Im}(\Omega)\leq 0$. Hence, the functions
\begin{equation}
I(\Omega,\mathbf{K})\equiv\int{d^3\mathbf{V}\frac{G(\mathbf{V,K})}{\Omega-\sqrt{3}\mathbf{K\cdot V}}},
\end{equation}
$\varepsilon(\Omega,K)$, and $\zeta(\Omega,K_\parallel)$ that make up the function $\tilde\rho(\Omega,\mathbf{K})$, must also be analytically continued into the lower half-plane of complex $\Omega$, thus extending their definition to the whole complex $\Omega$ plane. With thus continued functions, the contributions to the inverse Laplace transform (\ref{eq:inv_Laplace}) are of three sources~\cite{Guernsey_1969}: 
\begin{enumerate}
\item Contributions from the singularities of $I(\Omega,\mathbf{K})$ in the lower half of complex $\Omega$ plane (defined solely by the initial perturbation $G(\mathbf{V,K})$); with some simplifying assumptions about the initial perturbation~\cite{Guernsey_1969} these contributions are damped in a few plasma periods and can be ignored. 
\item Contribution of singularities of $1/\varepsilon(\Omega,K)$ in the lower half of complex $\Omega$ plane, of two types: (i) residues at the poles of $1/\varepsilon(\Omega,K)$, which give the volume plasma oscillations~\cite{Guernsey_1969}, and (ii) integrals along branch cuts (if any) of $1/\varepsilon(\Omega,K)$ in the lower half-plane of complex $\Omega$, which can lead to non-exponential attenuation of the volume plasma oscillations~\cite{Hudson_1962,Krivitskii_Vladimirov_1991}.
\item Contribution of singularities of $1/\zeta(\Omega,K_\parallel)$ in the lower half of complex $\Omega$ plane, of two types: (i) residues at the poles of $1/\zeta(\Omega,K_\parallel)$, corresponding to the surface wave solutions of the initial value problem in the considered system~\cite{Guernsey_1969}, and (ii) integrals along branch cuts (if any) of $1/\zeta(\Omega,K_\parallel)$ in the lower half-plane of complex $\Omega$, which will be discussed elsewhere. 
\end{enumerate}
In this paper, we will only consider the surface wave solutions due to the contribution of the residues at the poles of $1/\zeta(\Omega,K_\parallel)$. The dispersion and damping properties of these surface wave solutions are defined by the dispersion relation for plasma surface waves~\cite{Guernsey_1969}
\begin{equation}
\zeta(\Omega,K_\parallel)=0, \label{eq:zeta=0}
\end{equation}
which in case of stable plasma (with velocity distribution of electrons having only one maximum) only has non-growing solutions $\Omega=\Omega(K_\parallel)\in\mathbb{C}$ with ${\rm Im}(\Omega)\leq 0$~\cite{Penrose_1960}.

\subsection{Weakly damped surface waves}
% \textcolor{blue}{\textit{Argue that only the continuation to ${\rm Im}(\Omega)=0$ is needed (as we are only interested in weakly damped solutions of the dispersion equation for SW, with $\Gamma\ll\Omega$).}}

Out of all complex-valued solutions of (\ref{eq:zeta=0}), $\Omega(K_\parallel)=\Omega_s(K_\parallel) + i\Gamma_s(K_\parallel)$ (here $\Omega_s\in\mathbb{R}$ is the dimensionless frequency, and $\Gamma_s\leq0$ is the dimensionless damping rate of the surface wave), only those corresponding to \textit{weakly damped} surface waves, with $|\Gamma_s(K_\parallel)/\Omega_s(K_\parallel)|\ll 1$, are of physical interest. Such solutions can be obtained by solving, instead of the actual dispersion equation (\ref{eq:zeta=0}), the following set of approximate equations (that follow from (\ref{eq:zeta=0}) for $|\Gamma_s(K_\parallel)/\Omega_s(K_\parallel)|\ll 1$):
\begin{eqnarray}
{\rm Re}\left[\zeta(\Omega_s,K_\parallel)\right] &=& 0,\text{ yielding }\Omega_s=\Omega_s(K_\parallel)\in\mathbb{R} , \label{eq:Omega_s}\\
\Gamma_s &=& -\left.\frac{{\rm Im}\left[\zeta(\Omega_s,K_\parallel)\right]}{\partial{\rm\ Re}\left[\zeta(\Omega_s,K_\parallel)\right]/\partial\Omega_s}\right|_{\Omega_s=\Omega_s(K_\parallel)}, \label{eq:Gamma_s}
\end{eqnarray}
which only involve $\zeta(\Omega,K_\parallel)$ as a function of \textit{real} $\Omega$. Thus, instead of performing analytic continuation of $\zeta(\Omega,K_\parallel)$ from ${\rm Im}(\Omega)>0$ to ${\rm Im}(\Omega)\leq0$, it suffices to only analytically continue $\zeta(\Omega,K_\parallel)$ from ${\rm Im}(\Omega)>0$ onto the real axis ${\rm Im}(\Omega)=0$ of the complex $\Omega$ plane, thus defining it for $\Omega\in\mathbb{R}$.
% This analytic continuation of $\zeta(\Omega,K_\parallel)$, and the procedure for evaluating $\zeta(\Omega,K_\parallel)$ for $\Omega,K_\parallel\in\mathbb{R}$, required for the numerical solution of Eqs~(\ref{eq:Omega_s})--(\ref{eq:Gamma_s}), are discussed in Appendix~\ref{sec:zeta}. 
This analytic continuation of $\zeta(\Omega,K_\parallel)$, required for the numerical solution of Eqs~(\ref{eq:Omega_s})--(\ref{eq:Gamma_s}), is discussed in Appendix~\ref{sec:zeta}.

\section{Results and Discussion}
% \textcolor{blue}{\textit{Present the resulting plots of dispersion and damping rate, obtained from numerically solving Eqs~(\ref{eq:Omega_s})--(\ref{eq:Gamma_s}).}}

The numerical solution of Eqs~(\ref{eq:Omega_s}) and (\ref{eq:Gamma_s}) for a given $K_\parallel>0$, with $\zeta(\Omega,K_\parallel)$ defined for $\Omega\in\mathbb{R}$ as discussed in Appendix~\ref{sec:zeta}, yields the dispersion $\Omega_s(K_\parallel)$ and damping $\Gamma_s(K_\parallel)$ of surface waves; below we discuss them in greater detail.

\subsection{Dispersion}
% \textcolor{blue}{\textit{Discuss the small and large $K_\parallel$ asymptotes of the dispersion curve.}}

\begin{figure}
\includegraphics[width=4.0in]{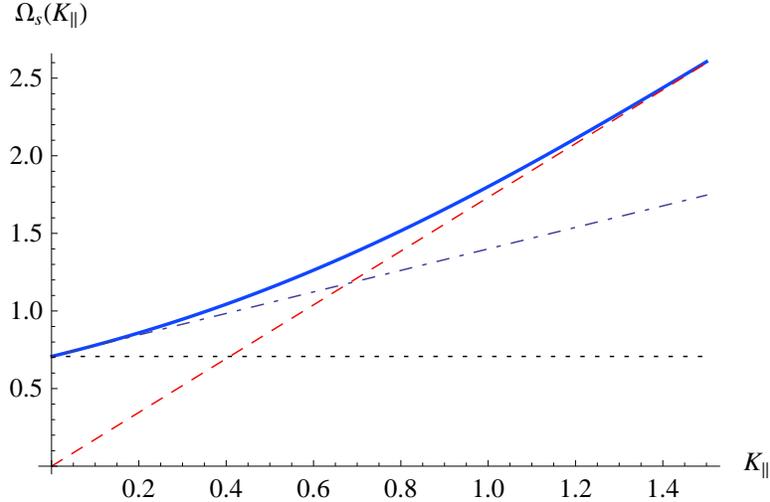}
\caption{\label{fig:dispersion} (Color online) The dispersion curve of surface waves $\Omega_s(K_\parallel)$ (solid blue line), and the asymptotes: the long-wave limit $1/\sqrt{2}$ (dotted line), the long-wave asymptote (\ref{eq:Omega_Kz<<1}) (dot-dashed blue line), and the short-wave asymptote (\ref{eq:dispersion_Kz>>1}) corresponding to zero sound propagating along the boundary (dashed red line).}
\end{figure}

The dispersion curve of surface waves $\Omega_s(K_\parallel)$ is shown in Fig.~\ref{fig:dispersion}. The SW frequency increases monotonically with $K_\parallel$, starting from the well-known long-wavelength limit of $\Omega_s=1/\sqrt{2}$ at $K_\parallel\rightarrow 0$ (which also corresponds to the cold-plasma limit of SW frequency), and increasing as 
\begin{equation}
\Omega_s(K_\parallel)\approx\frac{1}{\sqrt{2}}\left(1+0.95 K_\parallel\right)  \label{eq:Omega_Kz<<1}
\end{equation}
for $K_\parallel\ll 1$. We note the discrepancy between (\ref{eq:Omega_Kz<<1}) and the corresponding small-$K_\parallel$ asymptote of Lazar~et~al.'s result (\ref{eq:Lazar_disp}) of Ref.~\cite{Lazar_etal_2007}: $\Omega_s(K_\parallel)\approx(1/\sqrt{2})(1+\sqrt{3/2} K_\parallel)\approx(1/\sqrt{2})(1+1.23 K_\parallel)$. This discrepancy can not be attributed to the effect of quantum recoil, which is ignored here but included in the QFT model of Ref.~\cite{Lazar_etal_2007}, as the quantum recoil only gives the higher-order contribution $\propto H^2 K_\parallel^3$ to the dispersion (\ref{eq:Lazar_disp}) at small $K_\parallel$. In fact, the mentioned discrepancy is due to the error in the coefficient of the classical pressure gradient term used in Ref.~\cite{Lazar_etal_2007}: instead of the three-dimensional pressure gradient term $(3/5)v_F^2\nabla(n_e/n_0)$~\cite{Melrose_Mushtaq_PoP_2009,Eliasson_Shukla_PhysScr_2008}, the one-dimensional term $v_F^2\nabla(n_e/n_0)$ was used. With this error corrected, the QFT model of Ref.~\cite{Lazar_etal_2007} yields the small-$K_\parallel$ asymptote of $\Omega_s(K_\parallel)=(1/\sqrt{2})(1+\sqrt{9/10} K_\parallel)$ that matches our asymptote (\ref{eq:Omega_Kz<<1}).
 
% At short wavelengths, $K_\parallel\gg 1$, the SW frequency approaches the line $\Omega=\sqrt{3}K_\parallel$ (see Appendix~\ref{sec:dispersion_Kz>>1}) that corresponds to the dispersion of \textit{volume} zero sound mode in degenerate Fermi gas~\cite{Gol'dman_1947,LL_9} propagating along the boundary, with $K=K_\parallel$. Thus for $K_\parallel\gg 1$ the SW frequency becomes practically indistinguishable from the corresponding frequency of the volume plasma wave propagating along the boundary.

At short wavelengths, $K_\parallel\gg 1$, the SW dispersion can be approximated as (see Appendix~\ref{sec:dispersion_Kz>>1})
\begin{equation}
\Omega_s \approx \sqrt{3} K_\parallel \left(1+2\exp\left[-2-4K_\parallel^2\right]\right), \label{eq:dispersion_Kz>>1}
\end{equation}
and quickly approaches the line $\Omega=\sqrt{3}K_\parallel$ that corresponds to the dispersion of \textit{volume} zero sound mode in degenerate Fermi gas~\cite{Gol'dman_1947,LL_9} propagating along the boundary, with $K=K_\parallel$. Note that for $K_\parallel\gg 1$ the SW frequency $\Omega_s$ becomes practically indistinguishable from the corresponding frequency $\Omega_v$ of the volume plasma wave propagating along the boundary (with $K=K_\parallel$)~\cite{Gol'dman_1947}:
\[
\Omega_v = \sqrt{3} K_\parallel \left(1+2\exp\left[-2-2K_\parallel^2\right]\right).
\]

\subsection{Damping}
% \textcolor{blue}{\textit{Discuss the maximum of damping at $K_\parallel\approx 0.4$. Physical reason?}}

\begin{figure}
\includegraphics[width=4.0in]{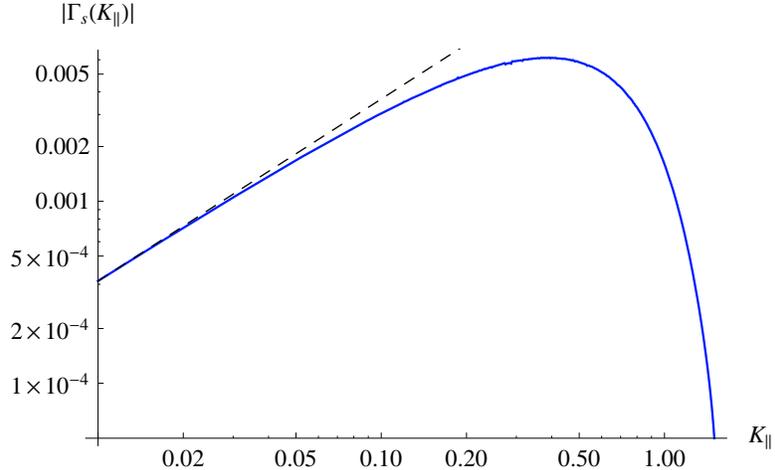}
\caption{\label{fig:damping} (Color online) The absolute value of surface wave damping rate $\left|\Gamma_s(K_\parallel)\right|$ (solid blue line), and the long-wave asymptote (\ref{eq:Gamma_Kz<<1}) (dashed line), as functions of $K_\parallel$.}
\end{figure}

The $K_\parallel$-dependence of the surface wave damping rate $\left|\Gamma_s(K_\parallel)\right|$ is shown in Fig.~\ref{fig:damping}. In the long-wave limit $K_\parallel\ll 1$, the damping rate increases linearly with $K_\parallel$ as
\begin{equation}
\left|\Gamma_s(K_\parallel)\right|\approx 2.1\sqrt{3}\cdot 10^{-2} K_\parallel,  \label{eq:Gamma_Kz<<1}
\end{equation}
similar to the damping rate of SW in plasma with Maxwellian electrons~\cite{Guernsey_1969,ABR_book}. However, as $K_\parallel$ increases, an important difference between SW damping in plasma with degenerate electrons and in plasma with Maxwellian electrons becomes obvious. In Maxwellian plasma, the SW damping rate increases monotonically with $K_\parallel$, quickly exceeding the SW frequency, so that the SW become strongly damped at short wavelengths~\cite{Guernsey_1969,ABR_book}. In degenerate electron plasma, however, the SW damping rate has a non-monotonic dependence on $K_\parallel$, as seen in Fig.~\ref{fig:damping}: at small $K_\parallel$, the damping rate increases almost linearly with $K_\parallel$, reaching a distinct maximum of $\left|\Gamma_s\right|\approx 6.2\cdot10^{-3}$ at $K_\parallel\approx 0.4$, and then \textit{decreases} monotonically with $K_\parallel$ for $K_\parallel>0.4$, approaching zero at large $K_\parallel$. 

% \textcolor{blue}{\textit{Point out that the waves remain weakly damped for all values of $K_\parallel$. This is the effect of Fermi degeneracy of electrons.}}

Most importantly, as seen from Fig.~\ref{fig:damping}, the SW damping rate in degenerate electron plasma remains small ($\left|\Gamma_s/\Omega_s\right|\ll 1$) at all values of $K_\parallel$, and hence the electrostatic surface waves in degenerate collisionless electron plasma are weakly damped at all wavelengths (yet the surface waves with $K_\parallel=k_\parallel\lambda_F\approx0.4$ are preferentially damped, as compared to other wavelengths).

The following note on the nature of damping of surface waves in collisionless plasma is in order. It is well known that in a Maxwellian plasma the collisionless (Landau) damping of surface plasma waves is not exponentially small and significantly exceeds that of the volume plasma waves in unbounded plasma, even at small value of $K_\parallel$ when their phase velocity greatly exceeds the electron thermal velocity~\cite{Guernsey_1969,ABR_book}. The reason is that, as seen from Eq.~(\ref{eq:zeta}), the surface wave field is essentially a result of plasma response to a collection of ``virtual'' plasma waves [i.e., having frequency $\Omega$ and wave vector $K$, but not having a dispersion relation $\Omega=\Omega(K)$, unlike the ``real'' plasma waves with dispersion relation $\Omega=\Omega(K)$ following from $\varepsilon(\Omega,K)=0$], with the same frequency $\Omega=\Omega_s$ but with different wave vectors $\mathbf{K}$ with absolute values $K=\sqrt{K_x^2+K_\parallel^2}$ ranging from $K=K_\parallel$ (for $K_x=0$) to $K\rightarrow\infty$ (for $K_x\rightarrow\infty$). Each of these virtual waves with its own $\mathbf{K}$ interacts resonantly with plasma electrons whose velocities along $\mathbf{K}$ is close to the phase velocity of the virtual wave (and such electrons always exist in Maxwellian plasma), and thus is subject to a non-zero Landau damping. As a result, the surface wave, consisting of both weakly damped virtual waves (with $\Omega<K$, where for Maxwellian plasma $K=k\lambda_D$, $\lambda_D=v_{Te}/\omega_p$ is the electron Debye length) and strongly damped virtual waves (with $\Omega>K$), interacts resonantly with a significant part of the plasma electron distribution, and hence is strongly damped, compared to the ``real'' plasma waves with the same wavelength.

In the degenerate plasma considered here, the difference in damping of surface and the ``real'' volume plasma waves is even more striking. Indeed, the ``real'' volume waves in such plasma, defined by the dispersion relation $\varepsilon(\Omega,K)=0$ with $\varepsilon(\Omega,K)$ given by Eq.~(\ref{eq:epsilon_degenerate}), are not subject to Landau damping at all, as their phase velocity exceeds the maximum possible velocity of degenerate plasma electrons, $\Omega>\sqrt{3}K$ (here $K=k\lambda_F$, $\lambda_F=v_F/\sqrt{3}\omega_p$ is the electron Fermi length), for all values of $K$~\cite{Gol'dman_1947}. The ``virtual'' plasma waves are, however, still subject to Landau damping by plasma electrons if their phase velocity is less than the electron Fermi velocity, $\Omega_s<\sqrt{3}K$, i.e., if their $K_x$ is large enough, $K_x>\sqrt{\Omega_s^2/3-K_\parallel^2}$; such virtual waves always give a contribution to the surface wave field, as $K_x$ in (\ref{eq:zeta}) attains arbitrarily large values. This also results in the finite, albeit small (yet infinitely large compared to the zero damping of the ``real'' volume plasma waves) Landau damping of surface waves in degenerate plasma.

\subsection{On applicability of the used approximations \label{sec:applicability}}

The collisionless mean-field approximation used here is justified for a system of weakly interacting particles, when two-particle correlations (and higher-order particle correlations as well) can be ignored. Physically, this corresponds to a system in which the collective effects dominate over the effects of particle collisions, which happens when the plasma coupling parameter $Q=U_{\rm int}/\epsilon_{\rm kin}$ is small (see Sec.~\ref{sec:model}). With $U_{\rm int}\sim e^2 n_0^{1/3}$ and $\epsilon_{\rm kin}\sim\epsilon_F$ for degenerate electrons, this implies that the collisionless mean-field approximation for degenerate electron plasma is justified when
\begin{equation}
Q\sim\frac{e^2 n_0^{1/3}}{\epsilon_F} \sim \left(\frac{\hbar\omega_p}{\epsilon_F}\right)^2 \ll 1. \label{eq:kappa<<1}
\end{equation}
% which is equivalent to the condition $e^2/\hbar v_F\ll 1$. On the other hand, the assumption of nonrelativistic plasma electrons, $v_F\ll c$, where $c$ is the speed of light, implies $e^2/\hbar v_F\gg e^2/\hbar c\approx 1/137$. This leads to a joint condition of validity of the nonrelativistic mean-field approximation:
% \begin{equation}
% 1/137\ll e^2/\hbar v_F \ll 1.	\label{eq:joint_condition}
% \end{equation}
Moreover, when the condition (\ref{eq:kappa<<1}) is satisfied, the effect of quantum recoil on the dispersion properties of electrostatic plasma waves is also negligible, at least for $K\lesssim1$~\cite{Manfredi_Haas_PRB_2001,Kuzelev_Rukhadze_UFN_2011}, and the only quantum effects come from the Fermi-Dirac statistics of plasma electrons. 
Thus the mean-field quasi-classical kinetic model (\ref{eq:Vlasov}), in which particle correlations and quantum recoil are neglected, is internally consistent under the condition (\ref{eq:kappa<<1}). We note that for degenerate electron plasma, the coupling parameter $Q\sim e^2 n_0^{1/3}/\epsilon_F$ scales as $n_0^{-1/3}$ and thus decreases as the plasma density $n_0$ increases; hence the condition (\ref{eq:kappa<<1}) is satisifed for sufficiently dense plasma, with $n_0\gg(e^2 m^*/\hbar^2)^3$, where $m^*$ is the effective electron mass in such plasma. However, as $n_0$ increases, the electron Fermi velocity increases as $n_0^{1/3}$ and may become comparable to the speed of light $c$, in which case the relativistic effects may become important; to avoid this, we also require that $n_0\ll (m^*c/\hbar)^3$. Thus the nonrelativistic mean-field quasi-classical approximation (\ref{eq:Vlasov}) is valid for quantum plasmas with degenerate electrons with densities
\begin{equation}
\left(\frac{e^2 m^*}{\hbar^2}\right)^3 \ll n_0 \ll \left(\frac{m^* c}{\hbar}\right)^3. \label{eq:joint_condition} 
\end{equation}
The condition (\ref{eq:joint_condition}) may be satisfied for some semiconductors, as pointed out in Ref.~\cite{Krivitskii_Vladimirov_1991}.

For electron gas in metals, however, the condition (\ref{eq:kappa<<1}) (as well as the first part of the condition (\ref{eq:joint_condition})) is not satisfied (in fact in metals $Q\gtrsim1$), which suggests that in metals the collisionless mean-field approximation (as well as the QFT model used in Ref.~\cite{Lazar_etal_2007}, that is derived from the mean-field kinetic model~\cite{Manfredi_Haas_PRB_2001}) is formally not justified, as the two- and higher-order particle correlations (due to their Coulomb and exchange interactions) become important. Instead, a kinetic model of moderately coupled plasma, accounting for particle correlations and quantum recoil, should be developed -- a difficult task, given the lack of a small parameter characterizing the electron interactions in metals. Nevertheless, there are reasons, discussed below, to believe that our idealized model can still work well (at least qualitatively) in metals, despite being formally unjustified there.

In the experimental work of Watanabe~\cite{Watanabe_1956}, the relation between energy loss and scattering angle of 25 keV electrons passing through thin metallic films was measured, and it was found, in particular, that the value of energy loss increases with the scattering angle of electrons. An empirical formula representing this relation for Be, Mg, and Al films was found to be in \textit{good agreement} with the corresponding theoretical formula that follows from the dispersion relation of electron plasma waves (excited by the energetic electrons traversing the metallic film), derived using the mean-field kinetic model of free electrons in metal, despite this model being formally unjustified in metals due to violation of (\ref{eq:kappa<<1}) there. Moreover, the effect of quantum recoil, although detectable, is small compared to the effect of Fermi-Dirac statistics in the whole range of measured scattering angles, especially at small scattering angles (corresponding to small values of plasma wave number). These facts can be perceived as an experimental evidence supporting the claim that the mean-field quasi-classical kinetic model can work well even in metals, despite there being $Q\sim 1$.

There is some theoretical evidence of this as well. In Ref.~\cite{Klim_Silin_UFN_1960}, the effect of electron correlations due to their exchange interaction was taken into account under the assumption of characteristic plasma perturbation wavelengths being large compared to the correlation length, using the Hartree-Fock approximation. In particular, it was found that the exchange correlations affect the spectrum of volume plasma waves by modifying their dispersion relation (in the limit of small $K$), which becomes:
\begin{equation}
\Omega^2 = 1 + \frac{9}{5} K^2 - \frac{9}{80}Q K^2,  \label{eq:spectrum_exchange}
\end{equation}
where the third term on the right is due to the exchange correlations, while the second term is due to the quantum statistics resulting in the Fermi-Dirac distribution (\ref{eq:f0}) of electron velocities in degenerate plasma. We see from (\ref{eq:spectrum_exchange}) that the term due to exchange correlations is small compared to the term due to quantum statistics even for $Q\sim 1$. This suggests that in metals the exchange correlations of electrons only lead to minor modification of plasma wave spectra, and neglecting them does not lead to a serious error, while greatly simplifying the model. As for the quantum recoil -- it modifies the spectrum (\ref{eq:spectrum_exchange}) by adding the higher-order term $(1/16)Q K^4$. Even for $Q\sim1$, the term due to quantum recoil remains small compared with the term due to quantum statistics for $K\lesssim 1$, and can be safely neglected.

Finally, it has beem pointed out in Ref.~\cite{Manfredi_Haas_PRB_2001} that in plasmas with low electron temperatures, $T_e/\epsilon_F\ll 1$ (e.g., for metals at room temperature), the effect of electron-electron Coulomb collisions is negligible. In fact, the typical electron-electron collision frequency $\nu_{ee}$ for metals at room temperature is of order $\nu_{ee}\sim 10^{10}\ s^{-1}$~\cite{Manfredi_Haas_PRB_2001}, which is many orders of magnitude smaller than the typical plasma oscillation frequency $\omega_p\sim 10^{16}\ s^{-1}$; hence the electron-electron collisions are not expected to play a significant role for processes occuring at the characteristic collective plasma time scale $\tau_p\sim\omega_p^{-1}$ (such as the surface waves studied here, with $\omega_s\sim\omega_p$), and can thus be neglected. Indeed, the dimensionless surface wave damping rate due to the electron-electron collisions, $\Gamma_s^{ee}=\gamma_s^{ee}/\omega_p$ with $\gamma_s^{ee}\sim\nu_{ee}$, is of the order $\Gamma_s^{ee}\sim10^{-6}$, which is much smaller than the characteristic collisionless damping rate $\Gamma_s\sim10^{-3}$ obtained in this work (see Fig.~\ref{fig:damping}), and thus can be safely neglected.

Beside colliding with each other, the electrons can also collide with the ions of metal lattice, yet these collisions were also neglected in our model. To justify this approximation, let us estimate the electron-ion collision frequency $\nu_{ei}$ in typical metals used in plasmonic applications, and compare the characteristic surface wave damping rate $\Gamma_s^{ei}$ associated with these collisions with the collisionless damping rate $\Gamma_s\sim10^{-3}$ obtained above. The experimentally measured electric resistivity of metals such as gold and aluminium at room temperature is of order $\rho\sim10^{-8}\text{ Ohm}\cdot\text{m}$\cite{Phys_Chem_book}. Using the definition of resistivity $\rho=E/j$, where $E$ is the applied electric field, and $j$ is the current density in metal (which is assumed to be entirely due to the free electrons of conductivity), and the Ohm's law $j=(\varepsilon_0\omega_p^2/\nu_{ei})E$ for electrons, where $\varepsilon_0=8.8542\cdot10^{-12}\text{ F m}^{-1}$ is the dielectric permittivity of vacuum, we have (in SI units):
\begin{equation}
\nu_{ei}\sim\varepsilon_0\omega_p^2\rho.  \label{eq:nu_ei}
\end{equation}
For metals, $\omega_p\sim10^{16}\text{ s}^{-1}$, and (\ref{eq:nu_ei}) yields $\nu_{ei}\sim10^{13}\text{ s}^{-1}$, which is small compared to the characteristic surface oscillation frequency, and thus is not expected to have a significant effect on the dispersion of surface waves. On the other hand, the dimensionless surface wave damping rate due to electron-ion collisions, $\Gamma_s^{ei}=\gamma_s^{ei}/\omega_p$ with $\gamma_s^{ei}\sim\nu_{ei}$, is of order $\Gamma_s^{ei}\sim\nu_{ei}/\omega_p\sim 10^{-3}$, which is comparable to the collisionless damping rate $\Gamma_s\sim 10^{-3}$ obtained here (recall that the maximum collisionless damping rate at $K_\parallel\approx0.4$ is $\Gamma_s\approx6.2\cdot10^{-3}$). Thus the electron-ion collisions, neglected in our model, may lead to additional collisional damping of surface waves that is comparable to the collisionless damping obtained from the collisionless kinetic model.

The above discussion suggests that the quasi-classical mean-field kinetic model used here is adequate for describing surface plasma waves even in moderately coupled plasmas with $Q\sim1$ such as the electron plasma in metals. To verify this, as well as to assess the qualitative and quantitative importance (or otherwise) of the effects associated with moderate values of the coupling parameter $Q$, a comparison of the model's predictions with experiments in various quantum plasmas, including the electron plasma in metals, is needed. 

\section{Conclusion}
In this paper, electrostatic surface waves in semi-bounded plasma with degenerate electrons were studied using the nonrelativistic collisionless mean-field kinetic model. The dispersion relation for the waves is obtained from the initial value problem, and its solution corresponding to weakly damped surface waves is presented, yielding dispersion and collisionless damping of the waves for an arbitrary wave number $K_\parallel$. It is shown that the surface waves in the semi-bounded plasma with degenerate electrons are weakly damped at all wavelengths, and their damping rate exhibits nonmonotonic dependence on $K_\parallel$, linearly increasing with $K_\parallel$ at $K_\parallel\ll1$, then reaching maximum at $K_\parallel\approx0.4$, then falling off rapidly to zero as $K_\parallel$ increases. This is in contrast with the strong damping of surface waves in semi-bounded plasma with Maxwellian electrons, and is the consequence of the effect of quantum statistics (leading to Fermi-Dirac velocity distribution) for plasma electrons. This work, using the more general kinetic model, extends the results of Ref.~\cite{Lazar_etal_2007} obtained using the quantum fluid theory, in two ways: (i) the range of the dispersion relation $\Omega_s(K_\parallel)$ is extended from $K_\parallel\ll1$ in Ref.~\cite{Lazar_etal_2007} to $K\lesssim 1$ in this paper, and (ii) the collisionless damping, absent in the model used in Ref.~\cite{Lazar_etal_2007}, is obtained and discussed here.  

\acknowledgments{This work was supported by the Australian Research Council. R.K. acknowledges the receipt of a Professor Harry Messel Research Fellowship funded by the Science Foundation for Physics within the University of Sydney.}

\appendix
\section{Analytic continuation of $\zeta(\Omega,K_\parallel)$ \label{sec:zeta}}
% \textcolor{blue}{\textit{Discuss the procedure for analytic continuation, and the behavior of the two roots of $\varepsilon(\Omega,K)=0$ in the complex $K_x$ plane in the process of decreasing ${\rm Im}(\Omega)$ to zero. The direction of the corresponding ``motion'' of the roots in $K_x$ plane crossing the real axis establishes the rule for pole bypassing in complex $K_x$ plane in the definition of $\zeta(\Omega,K_\parallel)$: the roots $K_x^r$ of $\varepsilon(\Omega,K)=0$ with ${\rm Re}(K_x^r)<0$ are bypassed from above, while the roots with ${\rm Re}(K_x^r)>0$ are bypassed from below, for ${\rm Re}(\Omega)>0$ (for ${\rm Re}(\Omega)<0$ the rule changes to the ``mirror reflection'' of itself).}}

To perform the required analytic continuation of $\zeta(\Omega,K_\parallel)$ onto the real axis of the complex $\Omega$ plane, we start from the definition (\ref{eq:zeta}) of $\zeta(\Omega,K_\parallel)$, with $\varepsilon(\Omega,K)$ given by (\ref{eq:epsilon_degenerate}), for ${\rm Im}(\Omega)>0$ (where $\zeta(\Omega,K_\parallel)$ is an analytic function of $\Omega$), and then reduce ${\rm Im}(\Omega)$ down to zero, taking the limit ${\rm Im}(\Omega)\rightarrow0+$ while ensuring that the analyticity of $\zeta(\Omega,K_\parallel)$ is preserved in the process. For ${\rm Im}(\Omega)>0$, $\zeta(\Omega,K_\parallel)$ is defined in terms of the integral 
\begin{equation}
\int_{-\infty}^{+\infty}{\frac{dK_x}{K^2\ \varepsilon(\Omega,K)}},  \label{eq:integral_Kx}
\end{equation}
where $K_\parallel>0$, $K_x\in\mathbb{R}$, $K=\sqrt{K_x^2+K_\parallel^2}$, and $\varepsilon(\Omega,K)$ is given by Eq.~(\ref{eq:epsilon_degenerate}). The function $\left[K^2\ \varepsilon(\Omega,K)\right]^{-1}$ under the integral in (\ref{eq:integral_Kx}) is an elementary function of $K=\sqrt{K_x^2+K_\parallel^2}$, which in turn is an elementary function of $K_x$. Thus the function $\left[K^2\ \varepsilon(\Omega,K)\right]^{-1}$ can be extended to complex $K_x$ plane by analytic continuation from the real axis ${\rm Im}(K_x)=0$ of the complex $K_x$ plane, which is achieved by taking the principal branches of the complex square root function $K=\sqrt{K_x^2+K_\parallel^2}$ and of the complex logarithm function $\ln[(\Omega+\sqrt{3}K)/(\Omega-\sqrt{3}K)]$ in (\ref{eq:epsilon_degenerate}), considered as functions of complex $K_x$. The resulting function $\left[K^2\ \varepsilon(\Omega,K)\right]^{-1}$ of complex $K_x$ has the following singularities in the complex $K_x$ plane:
\begin{enumerate}
\item Branch cut of the complex square root $\sqrt{K_x^2+K_\parallel^2}$, taken along the negative real axis of the argument $K_x^2+K_\parallel^2$. This branch cut maps into two branch cuts of $\left[K^2\ \varepsilon(\Omega,K)\right]^{-1}$ in the complex $K_x$ plane, given by two parametric equations:
\begin{equation}
K_x = \pm i \sqrt{K_\parallel^2 + \tau},\ \text{with }K_\parallel>0,\ \tau\in[0,+\infty). \label{eq:log_cuts}
\end{equation}  
\item Branch cut of the complex logarithm in (\ref{eq:epsilon_degenerate}), taken along the negative real axis of the argument $(\Omega+\sqrt{3}K)/(\Omega-\sqrt{3}K)$. This branch cut maps into two branch cuts of $\left[K^2\ \varepsilon(\Omega,K)\right]^{-1}$ in the complex $K_x$ plane, given by two parametric equations:
\begin{equation}
K_x = \pm i \sqrt{\frac{\Omega^2}{3}\left(\frac{\tau+1}{\tau-1}\right)^2 - K_\parallel^2},\ \text{with }\Omega\in\mathbb{C},\ K_\parallel>0,\ \tau\in[0,+\infty).
\end{equation}
\item Two poles $K_x=\pm i K_\parallel$ ($K_\parallel>0$) at the roots of $K^2=0$, lying symmetrically above and below the real axis of the complex $K_x$ plane.
\item Two poles $\pm K_x^{r}\in\mathbb{C}$ at the roots of $\varepsilon(\Omega,K)=0$. Note that for any $K\in\mathbb{R}$, $\varepsilon(\Omega,K)=0$ does not have roots with ${\rm Im}(\Omega)>0$, if the plasma equilibrium is stable~\cite{Penrose_1960}, which is the case considered here; therefore, for any ${\rm Im}(\Omega)>0$ the poles $\pm K_x^{r}$ are located \textit{away} from the real axis of the complex $K_x$ plane, and thus do not lie on the integration contour in (\ref{eq:integral_Kx}).
\end{enumerate} 
%%%%%%%%%%%%%%%%%%%%%%%%%%%%%%%%%%%%%%%%%%%%%%%%%%%%%%%%%%%%%%%%%%%%%%%%%%%%%%%%%%%%%%%%%%%%%%%%%%%
\begin{figure}
\includegraphics[width=3.0in]{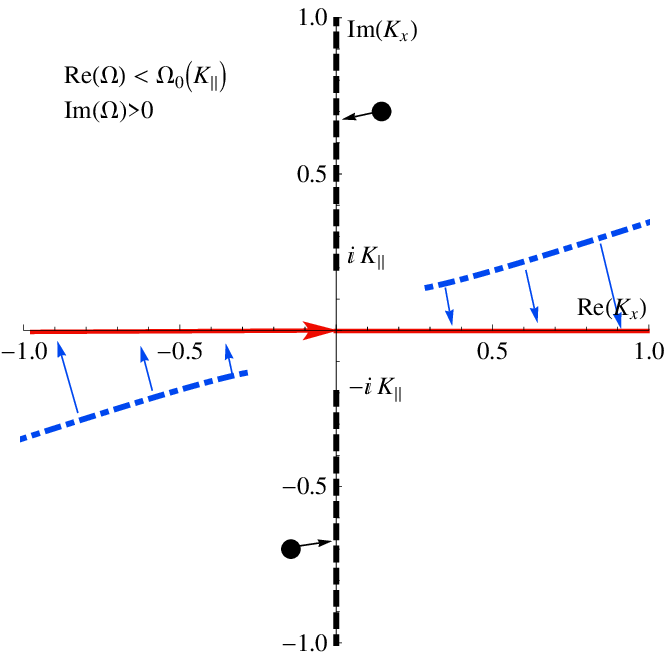}
\includegraphics[width=3.0in]{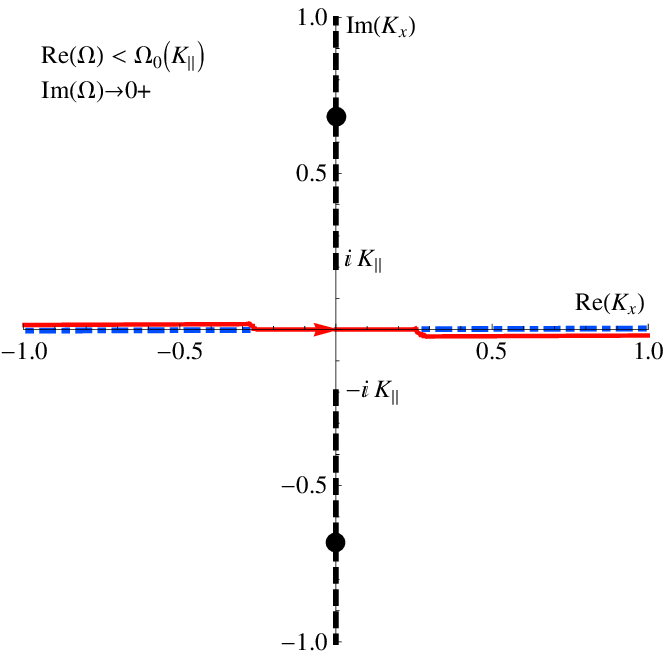}
\caption{\label{fig:Kx_plane_Omegar<Omega0} (Color online) Singularities (branch cuts and poles) of the function $\left[K^2\ \varepsilon(\Omega,K)\right]^{-1}$ in the complex $K_x$ plane, and their modification from ${\rm Im}(\Omega)>0$ (left panel) to the limit ${\rm Im}(\Omega)\rightarrow0+$ (right panel), for $0<{\rm Re}(\Omega)<\Omega_0(K_\parallel)$, with $\Omega_0(K_\parallel)$ defined from (\ref{eq:Omega0}). The branch cuts of the square root and the logarithm are shown with the black dashed lines and the blue dot-dashed lines, respectively. The poles $\pm K_x^{r}$ at the roots of $\varepsilon(\Omega,K)=0$ are shown with the filled circles. The arrows show the direction of motion of the singularities in the process of ${\rm Im}(\Omega)\rightarrow0+$. The contour of integration over $K_x$ in (\ref{eq:zeta}) is shown with the solid red line, with the arrow showing the direction of the contour.}
\end{figure}
\begin{figure}
\includegraphics[width=3.0in]{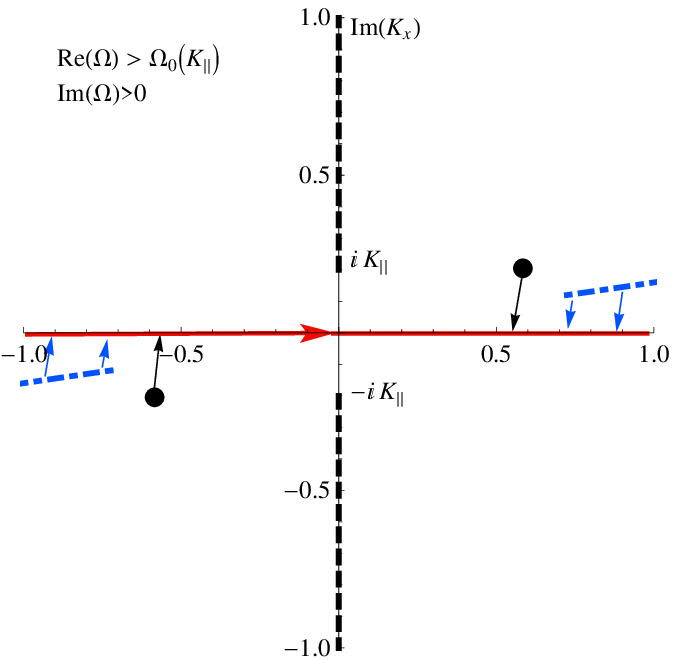}
\includegraphics[width=3.0in]{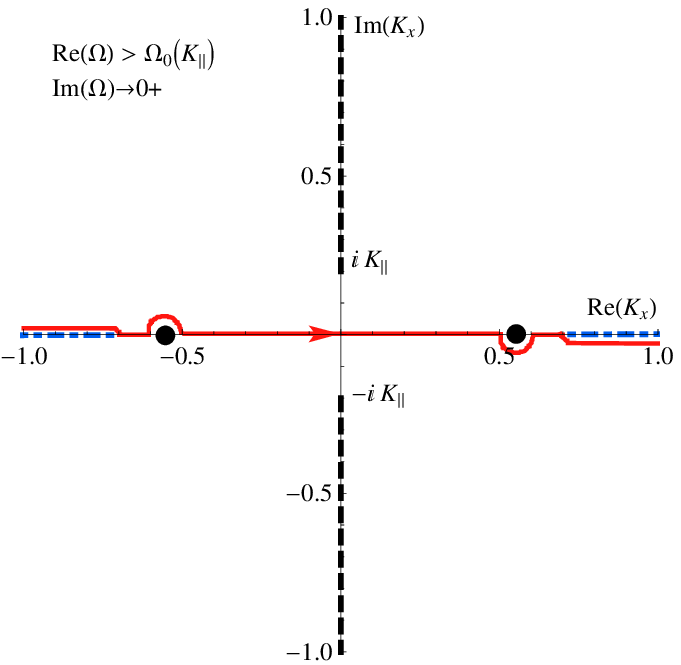}
\caption{\label{fig:Kx_plane_Omegar>Omega0} (Color online) Same as Fig.~\ref{fig:Kx_plane_Omegar<Omega0}, but for ${\rm Re}(\Omega)>\Omega_0(K_\parallel)>0$. The branch cuts of the square root and the logarithm are shown with the black dashed lines and the blue dot-dashed lines, respectively. The poles $\pm K_x^{r}$ at the roots of $\varepsilon(\Omega,K)=0$ are shown with the filled circles. The arrows show the direction of motion of the singularities in the process of ${\rm Im}(\Omega)\rightarrow0+$. The contour of integration over $K_x$ in (\ref{eq:zeta}) is shown with the solid red line, with the arrow showing the direction of the contour.}
\end{figure}
\begin{figure}
\includegraphics[width=3.0in]{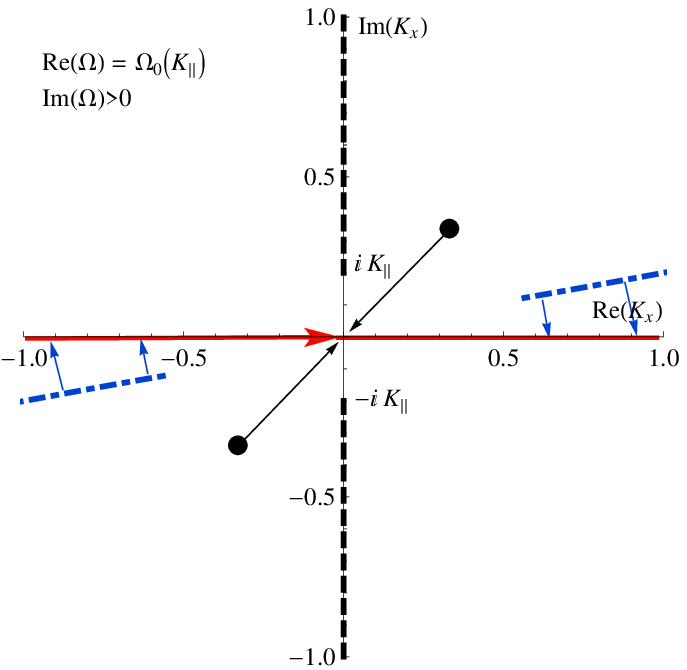}
\includegraphics[width=3.0in]{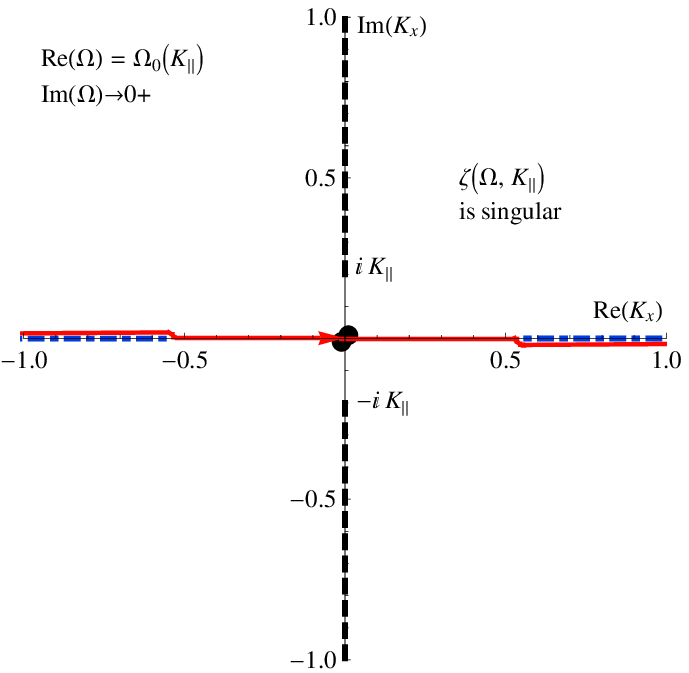}
\caption{\label{fig:Kx_plane_Omegar=Omega0} (Color online) Same as Fig.~\ref{fig:Kx_plane_Omegar<Omega0}, but for ${\rm Re}(\Omega)=\Omega_0(K_\parallel)>0$. The branch cuts of the square root and the logarithm are shown with the black dashed lines and the blue dot-dashed lines, respectively. The poles $\pm K_x^{r}$ at the roots of $\varepsilon(\Omega,K)=0$ are shown with the filled circles. The arrows show the direction of motion of the singularities in the process of ${\rm Im}(\Omega)\rightarrow0+$. The contour of integration over $K_x$ in (\ref{eq:zeta}) is shown with the solid red line, with the arrow showing the direction of the contour; for $\Omega=\pm\Omega_0(K_\parallel)$, this contour is ``squeezed'' between the poles $\pm K_x^{r}$, leading to a singularity in $\zeta(\Omega,K_\parallel)$ for ${\rm Im}(\Omega)\rightarrow0+$, when the poles $-K_x^{r}$ and $K_x^{r}$ merge.}
\end{figure}
%%%%%%%%%%%%%%%%%%%%%%%%%%%%%%%%%%%%%%%%%%%%%%%%%%%%%%%%%%%%%%%%%%%%%%%%%%%%%%%%%%%%%%%%%%%%%%%%%%%

The location of these singularities in the complex $K_x$ plane is shown in Figs~\ref{fig:Kx_plane_Omegar<Omega0}--\ref{fig:Kx_plane_Omegar=Omega0}. For ${\rm Im}(\Omega)>0$ (the left panels of Figs~\ref{fig:Kx_plane_Omegar<Omega0}--\ref{fig:Kx_plane_Omegar=Omega0}), none of the singularities of $\left[K^2\ \varepsilon(\Omega,K)\right]^{-1}$ lie on, or intersect with, the real axis ${\rm Im}(K_x)=0$ of the complex $K_x$ plane, along which the integration in (\ref{eq:integral_Kx}) is carried out; the function (\ref{eq:integral_Kx}) and, consequently, the function $\zeta(\Omega,K_\parallel)$ are thus analytic functions of $\Omega$ for ${\rm Im}(\Omega)>0$. However, as ${\rm Im}(\Omega)\rightarrow0+$ in the process of analytic continuation of $\zeta(\Omega,K_\parallel)$ to the real axis of complex $\Omega$ plane, some of the singularities of $\left[K^2\ \varepsilon(\Omega,K)\right]^{-1}$ move about in the complex $K_x$ plane as shown in Figs~\ref{fig:Kx_plane_Omegar<Omega0}--\ref{fig:Kx_plane_Omegar=Omega0}, and may collapse onto, or cross with the real axis, thus requiring deformation of the integration contour in (\ref{eq:integral_Kx}) to avoid crossing these singularities and to preserve analyticity of $\zeta(\Omega,K_\parallel)$. Below we consider three cases: (i) $|{\rm Re}(\Omega)|<|\Omega_0(K_\parallel)|$, (ii) $|{\rm Re}(\Omega)|>|\Omega_0(K_\parallel)|$, and (iii) ${\rm Re}(\Omega)=\pm\Omega_0(K_\parallel)$, where $\pm\Omega_0(K_\parallel)\in\mathbb{R}$ are found from the equation
\begin{equation}
\left.\varepsilon(\Omega,K)\right|_{K_x=0}=\varepsilon(\Omega,K_\parallel)=0,\ \text{for }\Omega,K_\parallel\in\mathbb{R}. \label{eq:Omega0}
\end{equation}

For $|{\rm Re}(\Omega)|<|\Omega_0(K_\parallel)|$, the logarithm branch cuts (\ref{eq:log_cuts}) collapse onto the real axis, while the poles $\pm K_x^{r}$ collapse onto the imaginary axis of the complex $K_x$ plane as ${\rm Im}(\Omega)\rightarrow0+$. The resulting integration contour in $K_x$ is shown in Fig.~\ref{fig:Kx_plane_Omegar<Omega0}.

For $|{\rm Re}(\Omega)|>|\Omega_0(K_\parallel)|$, both the logarithm branch cuts (\ref{eq:log_cuts}) and the poles $\pm K_x^{r}$ collapse onto the real axis of the complex $K_x$ plane as ${\rm Im}(\Omega)\rightarrow0+$. The resulting integration contour in $K_x$ is displaced to avoid the poles $\pm K_x^{r}$, as shown in Fig.~\ref{fig:Kx_plane_Omegar>Omega0}.

Finally, for ${\rm Re}(\Omega)=\pm\Omega_0(K_\parallel)$ (the boundary between the above two cases), the two poles $\pm K_x^{r}$ both collapse towards zero as ${\rm Im}(\Omega)$ decreases, as shown in Fig.~\ref{fig:Kx_plane_Omegar=Omega0}, ``squeezing'' the $K_x$ integration contour between them. In the limit ${\rm Im}(\Omega)\rightarrow0+$, the poles $\pm K_x^{r}$ merge at $K_x=0$, and the integration contour passes through both of them, resulting in $\zeta(\Omega,K_\parallel)$ being singular (non-analytic) at the point $\Omega=\pm\Omega_0(K_\parallel)$. 

\begin{figure}
\includegraphics[width=4.0in]{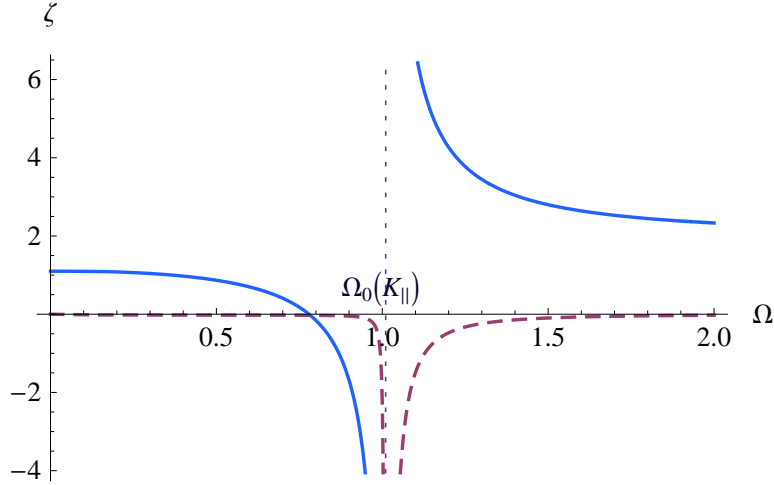}
\caption{\label{fig:zeta(Omega)} (Color online) The characteristic $\Omega$-dependence of $\zeta(\Omega,K_\parallel)$ analytically continued to ${\rm Im}(\Omega)=0$, for a fixed $K_\parallel=0.1$. ${\rm Re}[\zeta(\Omega,K_\parallel)]$ and ${\rm Im}[\zeta(\Omega,K_\parallel)]$ are shown with the solid blue and dashed purple lines, respectively. The frequency $\Omega_0(K_\parallel)$ defined by Eq.~(\ref{eq:Omega0}), at which $\zeta(\Omega,K_\parallel)$ is singular, is shown with the dotted vertical line.}
\end{figure}

The function $\zeta(\Omega,K_\parallel)$, defined for $\{\Omega,K_\parallel\}\in\mathbb{R}$ as described above, was calculated numerically by performing integration over $K_x$ along the appropriate one of the integration contours shown in Figs~\ref{fig:Kx_plane_Omegar<Omega0}--\ref{fig:Kx_plane_Omegar>Omega0}. The characteristic $\Omega$-dependence of thus defined $\zeta(\Omega,K_\parallel)$ is shown in Fig.~\ref{fig:zeta(Omega)}, for a fixed value of $K_\parallel>0$.

\section{Dispersion of surface waves for $K_\parallel\gg 1$ \label{sec:dispersion_Kz>>1}}
Consider the function $\zeta(\Omega,K_\parallel)$ defined by Eq.~(\ref{eq:zeta}), with $\varepsilon(\Omega,K)$ given by (\ref{eq:epsilon_degenerate}), whose definition is extended to $\Omega\in\mathbb{R}$ as described in Appendix~\ref{sec:zeta}. For any given $\Omega\in\mathbb{R}$, one can find $K_\parallel$ large enough so that $|\Omega|<|\Omega_0(K_\parallel)|$, where $\Omega_0(K_\parallel)$ is defined by Eq.~(\ref{eq:Omega0}) (this follows from the fact that $\Omega(K_\parallel)$ is a monotonically growing function of $K_\parallel$). Hence for sufficiently large $K_\parallel$, all poles of the function $1/K^2\varepsilon(\Omega,K)$ under the integral in (\ref{eq:zeta}) lie on the imaginary axis of the complex $K_x$ plane, symmetrically above and below $K_x=0$, as shown in Fig.~\ref{fig:Kx_plane_Omegar<Omega0}. Therefore, the ``projection'' of $1/K^2\varepsilon(\Omega,K)$ onto the integration contour along the real axis of complex $K_x$ plane is peaked at $K_x=0$ (the point on the integration contour closest to the poles) and decreases away from $K_x=0$ (as the distance from the poles increases). At $K_\parallel\gg 1$ the poles $K_x=\pm iK_\parallel$ of $1/K^2$ are located closer to the real axis of complex $K_x$ plane than the poles $K_x=\pm K_x^r$ of $1/\varepsilon(\Omega,K)$, hence the variation of the integrand in $\zeta(\Omega,K_\parallel)$ with $K_x$ is mainly defined by the variation of $1/K^2=1/(K_x^2+K_\parallel^2)$ along the real axis of complex $K_x$ plane. Neglecting the variation of $1/\varepsilon(\Omega,K)$ with $K_x$ (and using the peak value of $1/\varepsilon(\Omega,K)$ at $K_x=0$), we can thus approximate the integral over $K_x$ as 
\begin{equation}
\int_{-\infty}^{+\infty}{\frac{dK_x}{K^2\varepsilon(\Omega,K)}}\approx\int_{-\infty}^{+\infty}{\frac{dK_x}{(K_x^2+K_\parallel^2)\varepsilon(\Omega,K_\parallel)}} = \frac{\pi}{\varepsilon(\Omega,K_\parallel)K_\parallel},  \label{eq:int_approx}
\end{equation}
with $\varepsilon(\Omega,K_\parallel)$ defined by (\ref{eq:epsilon_degenerate}) with $K=K_\parallel$. The corresponding approximation for $\zeta(\Omega,K_\parallel)$ is then
\begin{equation}
\zeta(\Omega,K_\parallel)\approx \frac{1}{2}\left\{1+\left[1+\frac{1}{K_\parallel^2} - \frac{\Omega}{2\sqrt{3}K_\parallel^3}\ln\left(\frac{\Omega+\sqrt{3} K_\parallel}{\Omega-\sqrt{3} K_\parallel}\right)
\right]^{-1} \right\},\ \ K_\parallel\gg 1.  \label{eq:zeta_approx}
\end{equation} 

Substituting (\ref{eq:zeta_approx}) into the dispersion equation~(\ref{eq:Omega_s}) for weakly damped surface waves, we have
\begin{equation}
\frac{\Omega}{2\sqrt{3}K_\parallel^3}\ln\left|\frac{\Omega+\sqrt{3} K_\parallel}{\Omega-\sqrt{3} K_\parallel}\right| = 2 + \frac{1}{K_\parallel^2}. \label{eq:disp_tmp}
\end{equation}
Assuming the solution of (\ref{eq:disp_tmp}) to be of the form
\begin{equation}
\Omega = \sqrt{3}K_\parallel[1+\delta\Omega(K_\parallel)],\ \text{with }\lim_{K_\parallel\rightarrow\infty}|\delta\Omega(K_\parallel)|=0,     \label{eq:Omega_assumption}
\end{equation} 
we obtain for $\delta\Omega(K_\parallel)$:
\[
\delta\Omega(K_\parallel)\approx 2\exp\left[2-4 K_\parallel^2\right],
\]
which tends to zero at large $K_\parallel$, in agreement with the assumption (\ref{eq:Omega_assumption}). Thus we arrive at the approximation~(\ref{eq:dispersion_Kz>>1}) for the frequency of surface waves at $K_\parallel\gg1$.

%\bibliography{sw_paper}

\end{document}